# Precise and Fast Phase Wraps Reduction in Fringe Projection Profilometry

Minmin Wang[a], Guangliang Du[a], Canlin Zhou*[a], Shuchun Si[a], Zhenkun Lei**[b], XiaoLei Li [c],YanJie Li[d]

(a. School of Physics, Shandong University, Jinan , 250100, China
b. Department of engineering mechanics, Dalian University of Technology, Dalian, 116024, China
c.School of Mechanical Engineering, Hebei University of   Technology, Tianjin, 300130, China
d.School of civil engineering and architecture, Jinan University, Jinan, 2250020, China)
*Corresponding author: Tel: +8613256153609;     E-mail address: canlinzhou@sdu.edu.cn;
**Corresponding author: Tel: +8615841175236;   E-mail address: leizk@163.com

## Abstract

The number of phase wraps in 2D wrapped phase map can be completely eliminated, or greatly reduced by frequency shifting. But the wraps usually cannot be optimally reduced using the conventional fast Fourier transform (FFT) because the spectrum can be shifted only by an integer number in the frequency domain. In order to completely eliminate the phase wraps or achieve a significant phase wrap reduction, in this paper, we propose a fast and precise two-step method for phase wraps reduction, which uses the iterative local discrete Fourier transform (DFT) to determine the sub-pixel spectral peak location and the frequency shifting algorithm that operates in spatial domain to reduce the number of phase wraps. Firstly, an initial estimate of the frequency peak is obtain by FFT, then the sub-pixel spectral peak with high resolution is determined by iteratively upsampling the local DFT around the initial spectral peak location, further the non-integer frequency shifting in spatial domain is realized to eliminate or reduce the number of phase wraps. Finally, simulations and experiments are conducted to prove the validity of the proposed method. The results demonstrates the proposed method's superb computing efficiency, high resolution and overall performance.

## Keywords

Fourier transform; phase unwrapping; phase measuring profilometry (PMP); fringe projection

## 1. Introduction

Three-dimensional (3D) surface profiles can be obtained by using non-contact and high-accuracy fringe projection profilometry (FPP), which finds a wide variety of applications such as biological and medical imaging, computer vision and so on [1-3]. It is well known that the FPP heavily relies on the recovered phase, so that phase measuring profilometry (PMP) is generally a crucial process. The phase-shifting profilometry [4-5] and Fourier transform profilometry [6-7] can estimate the phase distribution by employing the arctangent of a quotient function. But the retrieved phase has $2\pi$ phase jumps. Therefore the phase unwrapping must be carried out in order to estimate the continuous phase map from the wrapped ones. For years, many different techniques have been developed to tackle phase unwrapping problems. These include branch-cut methods [8,9], quality guided approaches [10,11], minimum norm methods [12,13] and region-growing techniques [14].

Attempts have also been made at reducing the number of phase wraps in the phase map by estimating and eliminating the carrier, to reduce the complexity of phase unwrapping. Miguel et al. [15] proposed a algorithm to reduce the phase wraps by generating and subtracting the tilted plane from the entire phase map after estimating the modes of the first differences distribution in each axial direction. Feng et al. [16] proposed a carrier removal method based on principal component analysis, in which the phase of the carrier can be extracted from the first dominant component acquired. Chen et al. [17] developed a series expansion technique to remove the carrier phase component, where a least-squares method is used to estimate the unknown coefficients of the series in the carrier phase function. Quan et al. [18] compared and discussed several carrier-pase-component-removal techniques. Zhang et al. [19] presented a carrier removal method using Zernike polynomials fitting. Yue et al. [20] proposed a carrier removal method based on the analytical carrier phase description in phase measuring deflectometry. Du et al. [21] presented carrier-removal technique based on zero padding method. Yi et al. [22] proposed a carrier-removal method based on RBF interpolation.

Another way to reduce the number of phase wraps is by frequency shifting [23-24]. The number of phase wraps can be reduced or completely eliminated by transforming the image into the frequency domain, shifting the spectrum towards the origin, doing inverse Fourier transform and calculating the phase. But the use of the fast Fourier transform (FFT) limits the shift to integer values when using the method in Refs. [23-24]. However, the carrier frequency usually is not an integer but instead of a small fraction in practical applications. Thus, the phase wraps reduction is often not optimal, resulting in a seriously tilt error in the retrieved phase.

To increase the resolution and allow non-integer shift value in the frequency shifting, zero-padding approach [25,26] can be adopted by enlarging the array size of the original wrapped phase image and calculating its discrete Fourier transform. Though the method is accurate and effective, it has a heavy computational time and memory requirements.

In order to improve the calculation efficiency as well as simplify the procedures, here, we demonstrate a fast and precise two-step method for phase wraps reduction or elimination. Firstly, the frequency peak location is estimated to sub-pixel accuracy by iterative local discrete Fourier transform (DFT). Then the corresponding non-integer frequency shift is realized in spatial domain to eliminate or reduce the phase wraps. The sub-pixel accuracy technique here has previously appeared in the context of the image registration [27] and off-axis digital holograms [28] problem. This is the first time that it is used to reduce the number of phase wraps in Fringe Projection Profilometry. Different from the method in Ref. [28], the iterative local DFT algorithm is utilized [29] to reduce the amount of computed samples. Potential benefits of the proposed method are illustrated on both simulated and real fringe patterns.

The rest of the paper is organized as follows. Section 2 describes the theory and the computer simulation results. Section 3 shows experimental validation. Section 4 summarizes the full paper.

## 2. Theory and computer simulation

As for FPP, a number of sinusoidal fringes are projected onto the surface of the tested objects. Camera synchronously captures the distorted fringe patterns. Phase-shifting method or Fourier transform method is then used to demodulate the phase information, which is usually bounded in the range (-π, π]. In this paper, we process wrapped phase maps that have been produced using the four-step phase-shifting method. Besides, the method explained can be used equally as well to deal with wrapped phase maps that acquired using other methods to extract the phase.

The phase distribution of the Peaks function in Matlab is shown in Fig. 1(a) as a 3D surface, which consists of $N_A \times N_A$ pixels, $N_A$ =256. Four phase-shifted fringe patterns can be generated by using Eq. (1),

$$I_k(x, y) = a(x, y) + b(x, y)\cos[2\pi f_x x + 2\pi f_y y + \varphi(x, y) + \delta_k] \tag{1}$$

where $(x, y)$ are Cartesian coordinates, $a(x, y)$ the background intensity, $b(x, y)$ the intensity modulation amplitude, $\varphi(x, y)$ the phase to be retrieved. $\delta_k = k\pi/2$ is the phase step. $k$ = 0, 1, 2, 3, $k$ is image index. The parameter $f_x$ is the spatial carrier frequency along the x axis, while parameter $f_y$, the spatial carrier frequency along the y axis. Here, they are set to $f_x$ = 1/16 and $f_y$ = 0. One of the four fringe patterns is depicted in Fig. 1(b) as a gray scale range image.

The wrapped phase, $\varphi_w(x, y)$, is given by using Eq. (2), as shown in Fig. 1(c). It contains several wraps, and a phase unwrapping algorithm is needed to obtain the continuous phase.

$$\varphi_w(x, y) = \tan^{-1}\frac{I_3 - I_1}{I_0 - I_2} = W(2\pi f_x + 2\pi f_y + \varphi(x, y)) \tag{2}$$

*2.1 conventional method*

The phase wraps might be reduced to avoid the phase-unwrapping algorithm by using the Fourier transform method, proposed in [23], as follows. The Eq. (3) converts the wrapped phase map into the complex array $\varphi_{wc}(x, y)$,

$$\varphi_{wc}(x, y) = \cos[\varphi_w(x, y)] + i\sin[\varphi_w(x, y)] \tag{3}$$

where $i$ is equal to $\sqrt{-1}$. The 2D FFT is applied to $\varphi_{wc}(x, y)$ as shown in Eq. (4),

$$\phi_{wc}(u, v) = \mathcal{F}[\varphi_{wc}(x, y)] \tag{4}$$

where $\mathcal{F}[\cdot]$ is the 2D FFT operator, and the terms $u$ and $v$ are the vertical and horizontal frequencies respectively. The frequency spectrum is shifted towards the origin using the indices $u_o$ and $v_o$, which are determined by the spectral peak location and absolutely are integers. As for $f_x$ = 1/16 and $f_y$ = 0, the spectrum is shifted using $u_o$ = 256/16 = 16 and $v_o$ = 0, as shown in Fig. 1(d). Subsequently, the 2D inverse FFT of the shifted signal is computed as Eq. (5),

$$\psi(x,y) = \mathcal{F}^{-1}[\phi_{wc}(u-u_0, v-v_0)] \quad (5)$$

where $\mathcal{F}^{-1}[\cdot]$ is the 2D inverse FFT operator. Finally, the phase map can be extracted as follows,

$$\varphi_{wcs}(x,y) = \tan^{-1}\frac{\text{Im}[\psi(x,y)]}{\text{Re}[\psi(x,y)]} \quad (6)$$

where $\text{Im}[\cdot]$ represents the imaginary part, and $\text{Re}[\cdot]$ represents the real part of the complex array.

Following these procedure, the phase wraps in original phase map is eliminated or reduced. The absolute error between the theoretical phase and recovered one is calculated and shown in Fig. 1(e). However, the output of the conventional method, depends strongly on the location of the spectral peak. For instance, we change the spatial carrier frequency value into $f_x = 1/13$ and $f_y = 0$, as shown in Fig. 2(a). The spectral peak location should be $u_o = 256/13 \approx 19.6923$ and $v_o = 0$. As only integer location is normally possible, it leads to inaccurate results when the spectrum is shifted here using $u_o = 19$, $v_o = 0$ or $u_o = 20$, $v_o = 0$. The corresponding absolute errors are shown in Fig. 2(b) and 2(c), respectively.

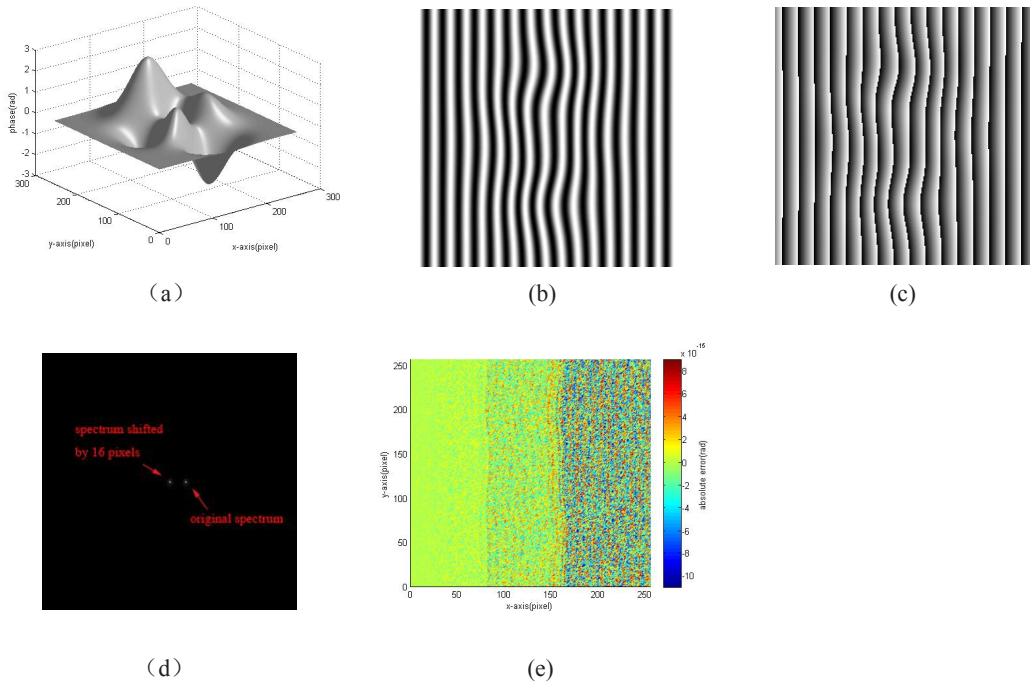

Figs. 1 Simulation results obtained using conventional method with a spatial carrier frequency of 1/16 fringes per pixel. (a) phase distribution of Peaks function; (b) One of the deformed fringe patterns; (c) Wrapped phase map; (d) The original spectrum and shifted one; (e) The absolute error of recovered phase.

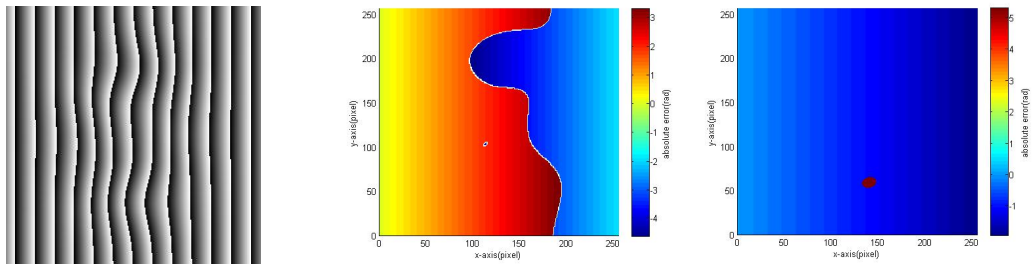

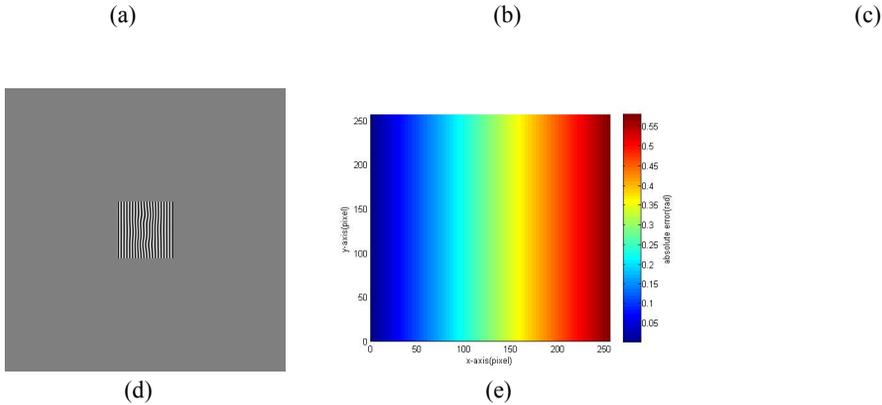

(a)　　　　　　　　　　　(b)　　　　　　　　　　　(c)

(d)　　　　　　　　　　　(e)

Figs. 2 Simulation results obtained using conventional method with a spatial carrier frequency of 1/13 fringes per pixel. (a) Wrapped phase map; (b)-(c) Absolute error of recovered phase using conventional method respectively shifting 19 steps and 20 steps in frequency domain; (d) Wrapped phase image zeros-padded to be fivefold the normal size; (e) The absolute error of recovered phase using zero-padding method.

*2.2 zero-padding method*

In an effort to reduce the error in the process of wraps reduction, the complex wrapped array $\varphi_{wc}(x,y)$ can be zero padded [25] for both the horizontal and vertical directions, then padded array is Fourier transformed. As for $f_x = 1/13$ and $f_y = 0$, Fig. 2(d) shows a wrapped phase image padded with zeros, the size of which is extended to be fivefold the normal size ($\alpha$=5) with the precision of 1/5 of the original frequency interval in the frequency domain. In this way, the distance of shifting frequency spectrum towards the origin can be corrected to one decimal place. The absolute error of the corresponding recovered phase is shown in Fig. 2(e). It is obvious that the zero-padding method achieves higher accuracy in reducing the phase wraps compared to conventional method (Fig. 2(b) or 2(c)). But the algorithmic complexity for a $\alpha$ fold zero-padded image is $O(\alpha^2 N_A^2 \log \alpha^2 N_A^2)$, meanwhile it has quite large of requirements for computational time and memory.

*2.3 proposed method*

Inspired by references [27-29], here, we propose a fast and precise two-step method for phase wraps reduction. The method mainly includes two steps: locating the frequency peak with high resolution and shifting the spectrum by the corresponding non-integer values.

To locate the frequency peak, we use 2D FFT to obtain an initial estimate of the frequency peak and then refine the frequency peak location to sub-pixel accuracy by upsampling the local DFT in a set of selected frequency coordinates around the initially estimated peak. Thus a high-resolution Fourier transform calculation is completed, and it is computationally much more efficient than a zero-padded FFT [22-23].

Assume that the DFT is calculated to an arbitrary fine sampling inside the area limited to the mask as a $N_B \times N_B$ array. The local high-resolution DFT may be calculated as a product of three matrices,

$$F(U,V) = \exp(\frac{-i2\pi UX^T}{N_A \alpha}) \bullet H(X,Y) \bullet \exp(\frac{-i2\pi YV^T}{N_A \alpha}) \qquad (7)$$

where the four vectors $\mathbf{U} = (u_0 ..... u_{N_B-1})^T$, $\mathbf{V} = (v_0 ..... v_{N_B-1})^T$, $\mathbf{X} = (x_0 ..... x_{N_A-1})^T$, $\mathbf{Y} = (y_0 ..... y_{N_A-1})^T$. And the vector elements are $x_k = y_k = (k - N_A/2) \times 1/N_A$ and $u_l = v_l = (l - N_A/2) \times m/N_A$, for $k = [0, ..., N_A - 1]$ and $l = [0, ..., N_B - 1]$. The vectors $\mathbf{U}$ and $\mathbf{V}$ are shifted to span the required sub-pixel spatial frequency locations centered on the initial spectral peak. If we calculate the local DFT for a 3 pixels neighborhood centered on the initial peak with an up-sampling factor of $\alpha = 100$ and $N_B = 3 \times 100 = 300$, the operation in Eq. (7) represents a multiplication of three matrices of sizes $300 \times N_A$, $N_A \times N_A$, and $N_A \times 300$, of which the algorithmic complexity is $O(\alpha N_A^2)$. Obviously, the reduction of the algorithmic complexity mainly depends on the up-sampling factor, and if the up-sampling factor $\alpha$ is far less than $N_A$, the algorithmic complexity of Eq. (7) will be significantly decreased compared to the zero-padding method.

In order to further decrease computing time and memory, we apply an iterative process [24]. For example, 1/100 of the original frequency interval can be achieved by two iterations using an up-sampling factor of $\alpha' = \alpha^{1/2} = 10$. Thus the algorithmic complexity for every iteration is reduced to be $O(\alpha' N_A^2)$. In the iterative process, the local DFT upsamples a $3 \times 3$ pixel region centered on the previous peak by an up-sampling factor of $\alpha' = 10$ and finds the current frequency peak in that array. At every iteration, the frequency precision is improved by 10 times. After two iterations, the frequency precision is improved by a factor of 100 compared to the initial resolution. More iterations can also be used in order to obtain the carrier frequency peak with higher accuracy.

After estimating the spectral peak location with high-resolution, the shifting distance (i.e. $u_o$ and $v_o$) of the frequency spectrum towards the origin can be accurately determined. Then we can reduce the number of phase wraps in spatial domain [25] as shown in Eq. (8).

$$\varphi_{wcs}(x,y) = F^{-1}[\Phi_{wc}(u+u_0, v+v_0)] = \varphi_{wc}(x,y) e^{(-j2\pi(u_o x/m + v_0 y/n))} \tag{8}$$

As the frequency shift is achieved through a simple multiply operation in spatial domain by the frequency shift property of 2D Fourier transform, the frequency shifting distance $u_0$ and $v_0$ can be the non-integers, not limited to be integers any more.

The main steps of the proposed algorithm can be summarized as follows:

(1) The wrapped phase is obtained using phase demodulation algorithm, such as the four-step phase-shifting method;

(2) The complex array is built according to Eq. (3);

(3) 2D FFT is used to obtain an initial estimate of the frequency peak location;

(4) Local DFT is applied to refine the spectral peak location to 1/10 of the original frequency interval;

(5) An iterative process is adopted to further increase the frequency precision by a factor of 100;

(6) The number of phase wraps is eliminated or reduced in spatial domain by Eq. (8).

Following the above procedure, we processed Fig. 2(a), and the result is shown in Figs. 3. The initial frequency peak is 20 pixel away from the origin confirmed by an FFT. For this location, Eq. (7) is utilized to calculate the frequency spectral value for a neighborhood centered on the initial peak and a new spectral peak (with higher resolution) is searched among the generated 30 × 30 spectral values, as shown in Fig. 3(a). In this way, the resolution of the frequency peak location was improved by ten times. In Fig. 3(b) we show distribution from a new 3 pixels neighborhood around the frequency peak location in Fig. 3(a). Two iterations will correct the peak location to two decimal places. The absolute error of the recovered phase for this case is shown in Fig. 3(c). Obviously, by applying the proposed method to the wrapped phase map, it acts to significantly reduce the absolute error.

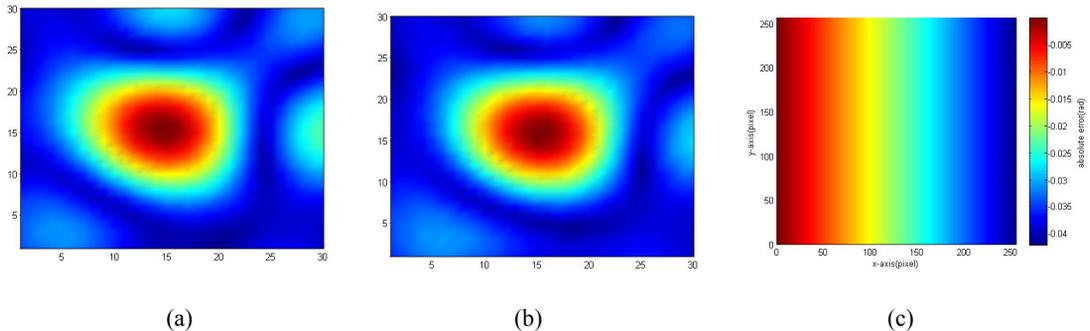

(a)  (b)  (c)

Figs. 3 Simulation results obtained using the proposed method with a spatial carrier frequency of 1/13 fringes per pixel. (a) DFT calculated in a 3 pixel neighborhood of the frequency peak with up-sampling factor of $\alpha'=10$; (b) DFT calculated in a new 3 pixels neighborhood of the frequency peak location in Fig. 3(b) with up-sampling factor of $\alpha'=10$; (c) The absolute error of recovered phase using the proposed method.

The root mean square error (RMSE) between the theoretical phase and the recovered one using the proposed method for the up-sampling factor of $\alpha'$ equal to 1, 5, 10, 15 and 20 are summarized in Table 1. We observe that the RMSE reduces with increasing up-sampling factor. In practical application, we may choose the suitable up-sampling factor in accord with different accurate demands,

Table 1. RMSE using the proposed method for the up-sampling factor equal to 1, 5, 10, 15 and 20

| up-sampling factor | 1 | 5 | 10 | 15 | 20 |
|---|---|---|---|---|---|
| RMSE(rad) | 1.1457 | 0.1084 | 0.0231 | 0.0143 | 0.0055 |

To compare the results of the conventional method, zero-padding method and proposed method, table 2 summarizes the accuracy (represented by RMSE of phase distribution) and the consuming time of the simulations mentioned above (Figs. 2 and 3). The spectrum is shifted using $u_o = 19$, $v_o = 0$ in the conventional method. The computational platform is a personal laptop with AMD Phenom II N660 @ 3.00GHz and 2.00GB RAM. We used MATLAB 2014a on the same computer to process the same images. As it can be seen from table 2, the proposed method significantly improve the accuracy with higher speed.

Table 2. Comparisons of RMSE and consuming time of three methods

|  | Conventional method | Zero-padding method | Proposed method |
|---|---|---|---|
| RMSE(rad) | 1.1457 | 0.3358 | 0.0231 |
| Consuming time(s) | 0.144 | 0.512 | 0.136 |

## 3. Experiments

In order to verify the feasibility of the proposed method, we developed a fringe projection measurement system, which consists of a DLP projector (Optoma DN344) and a

CCD camera (DH-SV401FM). The camera uses a 25 mm focal length mega-pixel lens (ComputarFAM2514-MP2) with a 688 × 582 pixels resolution, and a maximum frame rate of 50 frames/s. The 3D measurement software is programmed in MATLAB.

*4.1 simple face model with carrier frequency in one direction*

A face model with smooth shapes, of which the biggest height is about 3cm, was tested to verify the capability of the proposed method. One of the deformed fringe patterns captured by a CCD camera is shown in Fig. 4(a), as well as the corresponding wrapped phase obtained from the four-step phase-shifting method shown in Fig. 4(b). Firstly, zero-padding method was used to reduce the number of phase wraps. The array size of the original wrapped phase was enlarges to be fourfold the normal size by zero-padding (Fig. 4(c)) with the precision of 1/4 the original frequency interval. Then 2D FFT, frequency shifting and inverse FFT were applied to obtain the resultant phase, as shown in Fig. 4(c). This figure reveals that the $2\pi$ phase wraps have not been completely removed, and a phase unwrapping is still needed to obtain the continuous phase [9-11].

Then we adopted the proposed method. Initial spectral peak was estimated by an FFT, and iterative local DFT was used to determine the spectral peak location with higher resolution, as shown in Fig. 4(e) and 4(f). In this way, the spectral peak was found at 1/100 of the original frequency interval, namely the resolution of the frequency peak location was improved by hundred times. Finally, we reduced the number of phase wraps by Eq. (8). The resultant phase map is shown in Fig. 4(g). The time taken for the proposed method was 0.467 s, while the zero-padding method took 2.153 s. It demonstrates that the proposed method is of higher computing efficiency than the zero-padding method.

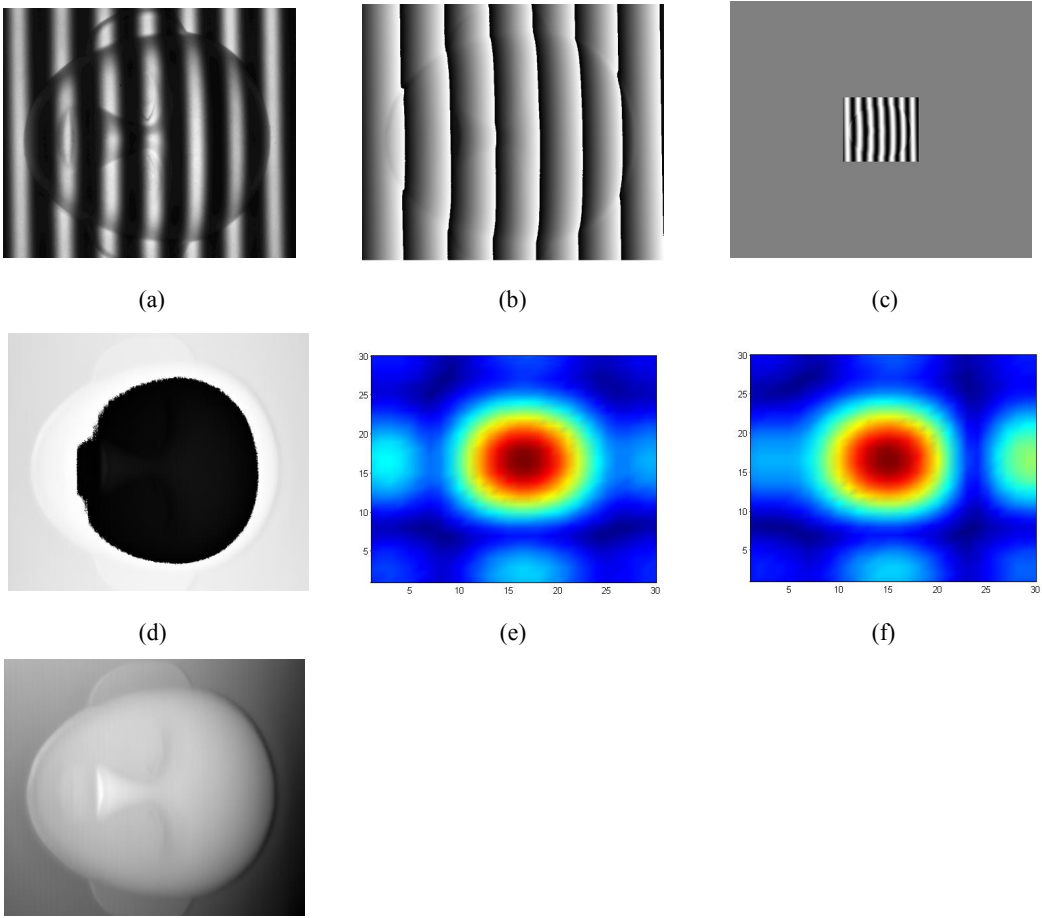

(a)          (b)          (c)

(d)          (e)          (f)

(g)

Figs. 4 Measurement result of the simple face model with carrier frequency in one direction. (a) One of the phase-shifted patterns; (b) Wrapped phase map; (c) Wrapped phase image zeros-padded to be fourfold the normal size; (d) Recovered phase using zero-padding method; (e) DFT calculated in a 3 pixel neighborhood of the frequency peak with up-sampling factor $α=10$; (f) DFT calculated in a new 3 pixels neighborhood of the frequency peak location in Fig. 4(e) with up-sampling factor $α=10$; (g) Recovered phase using the proposed method.

*4.2 complex model with carrier frequency in two directions*

In order to further test the performance of our proposed algorithm, a complex face model with the biggest height of about 5cm was measured. The fringe carrier (Fig. 5(a)) has corresponding components along x and y directions respectively. The procedure was similar to the previous experiment. Firstly, zero-padding method was used to reduce the number of phase wraps. The resultant phase is given in Fig. 5(d), it is obvious that the zero-padding method failed to completely eliminate phase wraps (the wrapped phase map is shown in Fig. 5(b)), even if the array size of the original wrapped phase was enlarges to be fourfold the normal size (Fig. 5(c)).

Then the proposed method was adopted. After determining the initial frequency peak location using FFT, the iterative local DFT was used to determine two refined spectral peaks, as shown in Fig. 5(e) and 5(f). We reduced the number of phase wraps by Eq. (8). The resultant phase is shown in Fig. 5(g). The time taken for the proposed method and the zero-padding method was 0.472 s and 2.236 s, respectively. Once again the proposed method provides an attractive solution for wraps reduction with high computation efficiency and accuracy.

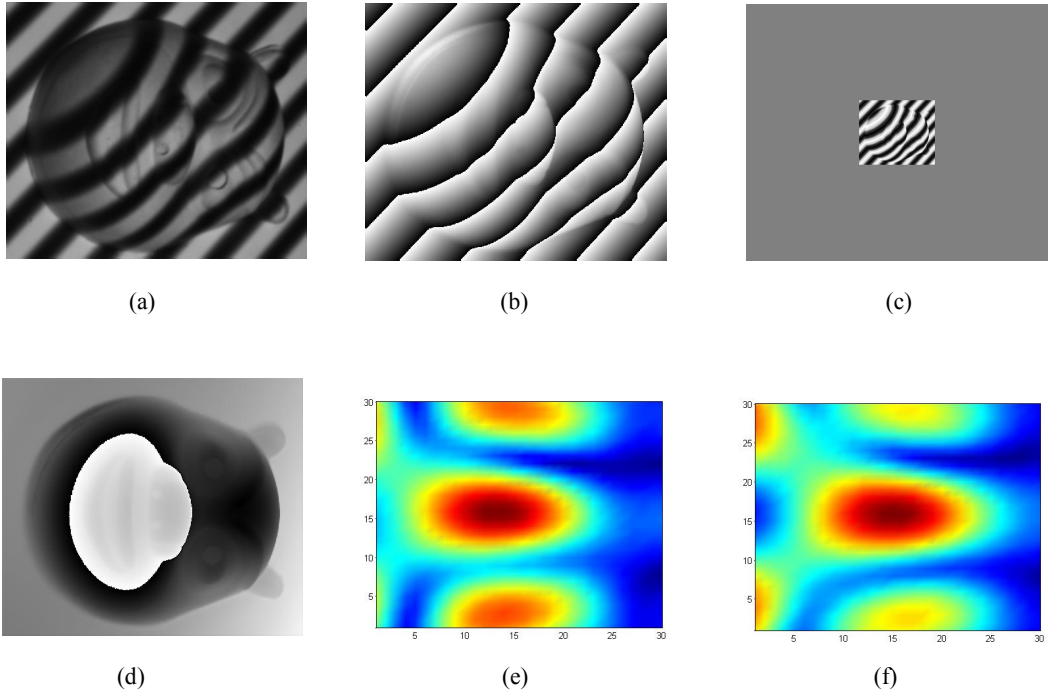

(a)  (b)  (c)

(d)  (e)  (f)

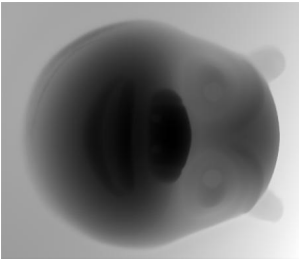
(g)

Figs. 5 Measurement result of the complex model with carrier frequency in two directions. (a) One of the phase-shifted patterns; (b) Wrapped phase map; (c) Wrapped phase image zeros-padded to be fourfold the normal size; (d) Recovered phase using zero-padding method; (e) DFT calculated in a 3 pixel neighborhood of the frequency peak with up-sampling factor $\alpha$=10; (f) DFT calculated in a new 3 pixels neighborhood of the frequency peak location in Fig. 5(e) with up-sampling factor $\alpha$=10; (g) Recovered phase using the proposed method.

## 4. Conclusion

In this paper, we propose a fast and precise two-step method for phase wraps reduction in fringe projection profilometry. An iterative local DFT is used to determine the spectral peak location with high resolution, and the frequency shift is achieved in spatial domain to eliminate or reduce the number of phase wraps. Based on simulations presented in Section 2 and the true fringe patterns analysis detailed in Section 3, the high computing efficiency and superb precision of the proposed method have been successfully demonstrated and confirmed. So the proposed method can simultaneously improve the operation speed and the accuracy of phase wraps reduction in profilometry.

**Acknowledgment**
This work was supported by the National Natural Science Foundation of China (NSFC) [Grant Nos: 11672162, 11302082 and 11472070]. The support is gratefully acknowledged.